\documentstyle[12pt]{article}

\def\ai{\'{\i}}
\def\eku{e^{2(K-U)}}
\def\eu{e^{2U}}
\def\-eu{e^{-2U}}
\def\om{\omega}
\def\2om{2\omega +3}
\def\ce{{\cal E}}
\def\8{8\pi G}
\def\pp{\varphi}
\def\4{4\pi G}
\def\16{16\pi G}

\def\be{\begin{equation}}
\def\ee{\end{equation}}
\def\ct{{\tilde E}_0}

\def\ct{{\cal T}}
\def\mt{energy-momentum tensor }
\begin{document}
\baselineskip.33in

\centerline{\large{\bf Cylindrical sources in full Einstein and Brans-Dicke  gravity}}

\bigskip

\centerline{ Andr\'es Arazi\footnote{Electronic mail: arazi@tandar.cnea.gov.ar} and Claudio Simeone\footnote{Electronic mail: simeone@tandar.cnea.gov.ar}}

\medskip
\centerline{\it Departamento de F\ai sica, Comisi\'on Nacional de Energ\ai a At\'omica}

\centerline{\it Av. del Libertador 8250, 1429 Buenos Aires, Argentina}

\vskip1cm

\noindent ABSTRACT

\bigskip

\noindent   It was shown by Hiscock that the energy-momentum tensor commonly used  to model local cosmic strings in linearized Einstein gravity can be extended and used in the full theory, obtaining a metric in the exterior of the source with the same deficit angle. Here we show that this tensor is an exception within a family 
for which a static solution  does not exist in full Einstein nor in Brans-Dicke  gravity.

\vskip1cm

{\it KEY WORDS:}  Static cylindrical solutions; cosmic string; Brans-Dicke.

\vskip1cm

{\it PACS numbers:} 98.80.Cq\ \ \ 04.50.+h\ \ \  11.27.+d

\newpage

\noindent{\bf 1. INTRODUCTION}

\bigskip

\noindent In a gauge theory, spontaneous symmetry breaking of a complex scalar field leads to cylindrical topological defects known as local cosmic strings [1,2]. The gravitational effects of such objects are of particular interest since they are considered as  possible ``seeds'' for galaxy formation  and  gravitational lenses. 
The metric around a local string was first calculated by Vilenkin [3] in the linear approximation of general relativity. Local strings are characterized by having an energy-momentum tensor whose only non null components are $T_t^{\, t}=T_z^{\, z}$. 
As linearized Einstein equations are formally analogous to Maxwell equations, the exterior solution does not depend on the radial distribution of the source. 
Hence, a  Dirac $\delta$ was used to approximate the radial distribution of the energy-momentum tensor for a cosmic string along the $z$ axis:
\be
{\tilde T}_\mu^{\, \nu}\equiv\delta (x)\delta (y)\int T_\mu^{\, \nu}(x,y)\,dx\,dy = \delta (x)\delta (y) \mbox{diag}(\mu,0,0,\mu),\ee
 where $\mu$ is the linear mass density. Under this assumptions, Vilenkin obtained a spacetime metric which is flat but with a deficit angle $\Delta \varphi =8\pi G\mu$, up to first order in $G\mu$ (in GUT strings $G\mu \sim 10^{-6}$). Since this metric has $g_{tt}=1$, i.e. the Newtonian potential is null, rest particles are not affected by the string. 

 Some years later, Hiscock [4], motivated by the possibility of theories which may lead to  values of $G\mu$ closer to one,                                                                                                                                                                                                                                     showed that Vilenkin's results are actually valid to  all orders in $G\mu$. As a source, he  considered a thick  cylinder of radius $a$ with uniform tension and linear mass density, whose tensor is
\be
 T_\mu^{\, \nu}(x,y)=\mbox{diag}(\mu,0,0,\mu){\theta(r-a)\over a^2}.
 \ee 
He solved {\em full} Einstein equations in the interior and matched the resulting static metric  with the vacuum solution for the exterior.

On the other hand, from the point of view of  structure formation it is important to determine whether an object interacts with rest particles. Vachaspati and Vilenkin  [5,6] obtained a metric  with non-null Newtonian potential 
considering a source whose tensor has  $T_z^{\, z}=\ct$ (effective tension) different from $T_t^{\, t}=\ce$ (energy per unit length). For this, they again  considered the  approximation of an infinitesimally thin ($\delta-$type) source and worked  within  linearized Einstein gravity. Similar results were obtained for such a source in linearized Brans-Dicke gravity [7].

In the present work we show that the case $T_t^{\, t}=T_z^{\, z}$ solved by Hiscock is an exception: thick sources with energy--momentum tensor  
\be
 T_\mu^{\, \nu}=\mbox{diag}({\ce},0,0,{\ct})   F(r)
\ee
  do not admit a static solution in  full Einstein nor in Brans--Dicke theories of  gravitation. In (3) $F(r)$ is any distribution function [8] whose integral  over the string transverse section is equal to unity.    
In this general case, we can obtain the static  metrics for the {\em exterior} by solving full Einstein and Brans--Dicke  vacuum equations for the most general metric with cylindrical symmetry. However,  we find that {\em static interior } solutions do not exist in either theories. 
 
\vskip1cm

\noindent{\bf 2. GENERAL RELATIVITY}

\bigskip

\noindent{\bf A. Weak field, $\delta$ source}

\medskip 

\noindent Vachaspati and Vilenkin [5] solved the linearized Einstein equations to obtain the metric in the exterior of an infinitesimally thin source described by the energy-momentum tensor
\be 
{\tilde T}_\mu^{\ \nu}=\mbox{diag} ({\ce} ,0,0,{\ct}) \delta(x)\delta(y).\ee
They found a solution which in cylindrical coordinates has the form
\begin{eqnarray}ds^2 & = & [1+4G(\ce-\ct)\ln(r/r_0)]\  dt^2\nonumber\\ & & \ \ \ \ \ \ \ \ -[1-4G(\ce+\ct)\ln(r/r_0) ]\ (dr^2+r^2 d\varphi^2)\nonumber\\ & &  \ \ \ \ \ \ \ \ \ \ \ \ \  \ \ \ -[1-4G(\ce-\ct)\ln(r/r_0)]\  dz^2
\end{eqnarray}
where $r_0$ is a constant of integration. As   $\ct\neq \ce$ we have $g_{00}\neq 1$ and, differing from the case $T_t^{\, t}=T_z^{\, z}$, there is an interaction with rest particles.
 
\bigskip

\noindent{\bf B. Full equations, finite cylindrical source}

\medskip

\noindent We shall start from the most general static metric with cylindrical symmetry  [9]:
\be
ds^2=\eku(dt^2-dr^2) -\-eu W^2 d\varphi^2-\eu dz^2 ,
\ee
where $K$, $U$ and $W$ are $r$--dependent functions. In terms of  these functions, 
the full Einstein equations for the \mt of equation (3) take the form:
\be
-{W''\over W}+K'{W'\over W} -{U'}^2 = \8 {\ce} F(r)\ \eku ,
\ee
\be
K'{W'\over W}-{U'}^2 =0,
\ee
\be
K''+{U'}^2=0,
\ee
\be
-{W''\over W}+2U''+2U'{W'\over W}-K''-{U'}^2=\8 \ct  F(r)\ \eku ,
\ee
where primes denote derivatives with respect to $r$. 
In the exterior of the source (\,$F(r)=0$\,) these equations lead to the 
 Weyl vacuum  metric which, with  our coordinates choice, has the form  
\be 
ds^2=r^{2d(d-1)}(dt^2-dr^2)-W_0^2r^{-2d}r^2d\varphi ^2-r^{2d}dz^2.
\ee
Since in a cylindrically symmetric problem the exterior solution is not independent of the interior metric, as it happens in a spherical problem,   the integration constants $W_0$ and $d$ should be determined by matching both metrics in the boundary. With $d=0$ or $d=2$ the metric (11) becomes Lorentz invariant in the $z$ direction; the case  $d=0$ is the one solved by Hiscock.

We shall show, however, that a static interior solution does not exist. From (9) and (10) we have
\be
-{W''\over W}+2U''+2U'{W'\over W}=\8 \ct\  F(r)\ \eku ,
\ee
and from (7) and (8)
\be
 -{W''\over W}=\8 {\ce}\ F(r) \eku,
\ee
so that
\be
U''+U'{W'\over W}={1\over 2}{W''\over W}\left({{\ce}-{\ct}\over{\ce}}\right).
\ee
The conservation equation [10]
\be
{T_\rho ^{\, \sigma}}_{;\sigma}={\partial\over\partial x^\sigma}\left(T_\rho ^{\ \sigma} \sqrt{-g} \right)-{1\over 2}\sqrt{-g}\ {\partial g_{\sigma\tau} \over\partial x^\rho}\ T^{\sigma\tau}=
0
\ee
 yields
\be
(K'-U'){\ce} +U'{\ct}=0.
\ee
Using this equation  we can write $K'=U'({\ce}-{\ct})/{\ce}$ and  $K''=U''({\ce}-{\ct})/{\ce}$, and then
from (8) and (9) we obtain
\be
\left(U''+ U'{W'\over W}\right)\left({{\ce}-{\ct}\over{\ce}}\right)=0.\ee
In the particular case  ${\ct}={\ce}$ these equations are compatible and yield   the interior metric found by Hiscock ($g_{\pp \pp}=-(a^2/ \8 \ce)\sin^2(\sqrt{\8 \ce}\,r/a)$~). However, for ${\ct}\neq {\ce}$ equations (14) and (17)  yield  $W''=0$. If so, equation (13) gives ${\ce}=0$, 
which means that there is no string. Hence  an  interior static solution cannot exist in the full theory. 

\vskip1cm

\noindent{\bf 3. BRANS-DICKE GRAVITY}

\bigskip

\noindent In the framework of present unified theories a scalar field should exist besides the metric of the spacetime. Scalar-tensor theories of gravitation would be important when studying the early universe, where it is supposed the coupling of the matter to the scalar field could be nonnegligible. Topological defects are produced in phase transitions in the early universe, so that it seems natural to study them in  a scalar-tensor theory of gravitation as that of Brans and Dicke [11,12,13].

In Brans-Dicke theory [14,15] matter and nongravitational fields generate a long-range scalar field $\phi$, which, together with them, acts as a source of gravitational field. The field $\phi$ is a solution of the equation 
\be
\phi_{;\sigma}^{\ ;\sigma}={1\over\sqrt{-g}}{\partial\over\partial x^\sigma}\left(\sqrt{-g}\ g^{\sigma\tau}{\partial\phi\over\partial x^\tau}\right)={8\pi T\over\2om}
\ee
where $T=\delta_\mu^{\ \nu} T_\nu^{\ \mu}$ and $\om$ is a dimensionless constant; the metric equations replacing those of General Relativity are 
\be
R_{\mu\nu}-{1\over 2}g_{\mu\nu}R=8\pi {T_{\mu\nu}\over\phi}+{\om\over\phi^2}\phi_{,\mu}\phi_{,\nu}-{\om\over 2\phi^2} g_{\mu\nu}\phi_{,\alpha}\phi^{,\alpha}+{1\over\phi}\phi_{,\mu;\nu}-{1\over\phi}g_{\mu\nu} \phi_{;\sigma}^{\ ;\sigma}.
\ee
\medskip

\noindent{\bf A. Weak field, $\delta$ source}

\medskip

\noindent In the linearized approximation the $\phi$ field   is expanded as $\phi\approx\phi_0+\xi= G^{-1}+\xi$ so that the equations for the metric and $\phi$  are 
\be R^{(1)} _{\mu\nu}=\8 \left(T_{\mu\nu}- {\om+1\over\2om}\eta_{\mu\nu} T\right)+G\xi_{,\mu,\nu},\ \ \ \ \ \ \ \ \ \Box\phi=\Box\xi={8\pi T\over\2om}.\ee 
 In the Brans-Dicke gauge ~$ (h_\mu^\nu-\delta_\mu^\nu\ h)_{,\mu}=G\xi_{,\nu}$ ~the perturbation $h_{\mu\nu}$ decouples from $\phi$ and the equations for the metric take the simple form [16]
\be\nabla^2 h_{\mu\nu} =\16 \left( T_{\mu\nu}-{\om+1\over\2om}\eta_{\mu\nu}T\right).\ee
Solving this equations for  the energy-momentum tensor  (4)  we obtain [7]
\begin{eqnarray}ds^2 & = &   \left[1+{8G\over\2om}[\ce(\om+2)-{\ct}(\om+1)]\ln(r/r_0)\right]\, dt^2\nonumber\\ &  &   \ \ \ \ \ \ \ \ \ -\left[1-8G(\ce+{\ct})\left({\om+1\over\2om}\right)\ln(r/r_0)\right]\, (dr^2+r^2 d\varphi^2)\nonumber\\ & & \ \ \ \ \ \ \ \ \ \ \ \ \  \ \ \ \  -\left[1-{8G\over\2om}[\ce(\om+1)-{\ct}(\om+2)]\ln(r/r_0)\right]\,  dz^2.
\end{eqnarray}
In the limit $\om\,\to\,\infty$ the metric (5) is recovered. If we write 
$$ ds^2=g_{00}\left( dt^2+\sum_{i=1} ^3 g^{00}g_{ii} (dx^i)^2\right)$$
and redefine the radial coordinate by 
$$ (1-8G\ce \ln(r/r_0))r^2= (1-8G\ce )\rho^2,\ \ \ \ \ \   (1-8G\ce \ln(r/r_0))dr^2\approx d\rho^2,$$
 we can put the metric in the form
\begin{eqnarray}
ds^2 & = & \left(1+{8G\over\2om}[\ce(\om+2)-{\ct}(\om+1)] \ln (\rho/\rho_0)\right)\times\nonumber\\ & &  \ \ \ \ \ \ \ \ \ \times  \left(dt^2-d\rho^2-(1-8G\ce)\rho^2 d\varphi^2    -[1-8G(\ce-{\ct}) \ln (\rho/\rho_0)]  dz^2\right).
\end{eqnarray}
In a plane perpendicular to the $z$ axis the metric is conformal to one with a deficit angle 
$\Delta = 8\pi G\ce $
 which does not depend on the Brans-Dicke constant $\om$. 

\bigskip

\noindent{\bf B. Full equations, finite cylindrical source}

\medskip

\noindent For the source (3), the Brans-Dicke equations (19) read  
\be
-{W''\over W} + K'{W'\over W} -{U'}^2 = 8\pi\  {{\ce}\over\phi} \ F(r)\ \eku +{\om{\phi'}^2\over 2\phi^2} - (K'-U'){\phi'\over\phi} + {1\over\phi} \left(\phi''+\phi' {W'\over W}\right)
\ee
\be
K'{W'\over W} - {U'}^2 = {\om{\phi'}^2\over 2\phi^2} -(K'-U'){\phi'\over\phi} - {W'\phi'\over W\phi}
\ee
\be
K''+{U'}^2=-{\om{\phi'}^2\over 2\phi^2}- U'{\phi'\over\phi}-{\phi''\over\phi}
\ee
\be
-{W''\over W} + 2U''+2U'{W'\over W}-K''-{U'}^2= 8\pi\ {{\ct} \over\phi}\  F(r)\ \eku +{\om{\phi'}^2\over 2\phi^2}+{1\over\phi} \left(\phi''+\phi' {W'\over W}\right)- U'{\phi'\over\phi}
\ee
and the equation (18) for the $\phi$ field takes the form 
\be
\phi''+\phi' {W'\over W}=-8\pi\ \left( {{\ce}+{\ct}\over\2om}\right)\ F(r)\ \eku~.
\ee
We shall first find the metric in the {\em exterior} of the source by solving these equations for vacuum, that is, with $F(r)=0$. For this case,  we inmediately see that $\phi=1/G$ is a particular solution of (28) which leads to the equations of general relativity.
To find the general solution we shall subtract equation (24) from   (25) to  get 
\be W\phi=ar+b.\ee
 Adding (25) and (26) and using (28) we obtain
$$K'=c{a\over W\phi}$$
while adding (26) and (27), with the use of (28) and (29), we get
$$U'=d{a\over W\phi}~.$$ 
This yields
$$U=d\ln \left| {ar+b \over e}\right|,\ \ \ \ \ \ \ K=c\ln \left| {ar+b \over f}\right| ,$$
\be W=g(ar+b)^n,\ \ \ \ \ \ \  \phi={1\over g}(ar+b)^{1-n},\ee
where $a\ldots g,n$ are integration constants; from (24) or (25)  the relation
$c=d(d+1-n)+{1\over 2}\om (1-n)^2+n(n-1)$ can be obtained. 
The resulting metric can therefore be put in the form
\be ds^2=r^{2d(d-n)+(\om+2n)(n-1)}(dt^2-dr^2)-W_0^2 r^{2(n-d)}d\varphi^2-r^{2d} dz^2.\ee
Choosing $n=1$ (which corresponds to $\phi=constant$) the Weyl metric of equation (11) is recovered.

Now let us study the possibility of obtaining a static solution for the {\em interior} of the source.
From (25), (26) and (28) we obtain
\be
K''+K'\left({W'\over W}+{\phi'\over \phi}\right)={8\pi\over \phi}\left( {{\ce}+{\ct}\over\2om}\right)\ F(r)\ \eku ,
\ee
and from (24) and (27)
\be
K''-2U''+(K'-2U')\left({W'\over W}+{\phi'\over \phi}\right)=8\pi\left({{\ce}-{\ct} \over\phi}\right)\  F(r)\ \eku . 
\ee
Now, using the  conservation equation (16), equation (33) can be put as 
\be
\left({{\ce}+{\ct}\over{\ce}-{\ct}} \right)\left[K''+K'\left({W'\over W}+{\phi'\over \phi}\right)\right]=-8\pi\left({{\ce}-{\ct} \over\phi}\right)\  F(r)\ \eku .  
\ee
Then, comparing this  with equation (32) we find that it should be
\be
-\left({{\ce}+{\ct}\over{\ce}-{\ct}} \right)^2=\2om .
\ee
Hence, for ${\ct}\neq {\ce}$ a static solution would be possible only if 
\be 
\om<-{3\over 2},
\ee
 which corresponds (see reference 14) to  $G<0$, that is, to a theory in which gravitation is  repulsive. Hence a static solution for the interior metric cannot exist in Brans-Dicke full theory.

\vskip1cm

\noindent{\bf 4. DISCUSSION}

\bigskip

\noindent Hiscock showed that the deficit angle obtained by Vilenkin [3] within the linear approximation of general relativity is correct to all orders in $G\mu$. For this, he used a thick cylinder as a source and found an exact  interior static solution which he matched with the exterior metric. Here we have shown that this procedure cannot be carried out with a more general tensor with  $T_t^{\, t}\neq T_z^{\, z}$: in this case there is no static interior solution.     

Tensors of this kind were considered by  Vachaspati and Vilenkin [5] when they studied the effect of wiggles propagating along a  string.  They used ${\tilde T}_\mu^{\ \nu}=\mbox{diag} ({\ce} ,0,0,{\ct}) \delta(x)\delta(y)$ for calculating the exterior metric within  linearized Einstein gravity.  The energy per unit length ${\ce}$ and the effective tension ${\ct}$ (fulfilling ${\ce}{\ct}=\mu^2$) are obtained by averaging over a distance and a time much greater than the typical wavelength and oscillating period of  the wiggles.   It may be thought, following Hiscock's idea, that a natural extension would be to use   $T_\mu^{\ \nu}=\mbox{diag} ({\ce} ,0,0,{\ct}) F(r)$ and to find a stationary interior solution; this solution would be the time average of the actual time-dependent metric (this implies neglecting gravitational radiation). In this picture, $F(r)$ would play the role of a spatial distribution obtained  as a time average of the radial position of the string; the interior of the source would be  the region defined by  $r$ less than the maximum amplitude of the wiggles. Regardless of the validity of this approximation,  it is clear from our analysis that the inmediate extension valid for local straight strings 
cannot be applied in the case of wiggly strings. To obtain a solution valid to all orders it may be necessary either to improve the approximations made in the  \mt or to consider the possibility of a time-dependent solution.

\vskip1cm

\noindent{\bf ACKNOWLEDGMENT}

\bigskip

\noindent We wish to thank F. D. Mazzitelli for reading the manuscript and  making helpful comments.

\newpage

\noindent{\bf REFERENCES}

\bigskip

\noindent 1. A. Vilenkin and E. P. S. Shellard, {\it Cosmic Strings and Other Topological Deffects}, Cambridge University Press, Cambridge (1994).

\noindent 2. A. Vilenkin, Phys. Rep. {\bf 121}, 263 (1985).

\noindent 3. A. Vilenkin, Phys. Rev. {\bf D 23}, 852 (1981).

\noindent 4. W. A. Hiscock, Phys. Rev. {\bf D 31}, 3288 (1985).

\noindent 5. T. Vachaspati and A. Vilenkin, Phys. Rev. Lett. {\bf 67}, 1057 (1991).

\noindent 6. T. Vachaspati, Phys. Rev. {\bf D 45}, 3487 (1992).

\noindent 7. A. Arazi and C. Simeone, submitted to Phys. Rev. {\bf D}.

\noindent 8. B. Linet, Gen. Rel. Grav. {\bf 17}, 1109 (1985).

\noindent 9. K. S. Thorne, Phys. Rev. {\bf 138}, 251 (1965).

\noindent 10. L. D. Landau and E. M. Lifshitz, {\it The Classical Theory of Fields}, Pergamon Press, Oxford (1975).

\noindent 11. A. A. Sen, N. Banerjee and A. Banerjee, Phys. Rev. {\bf D 56}, 3706 (1997).

\noindent 12. C. Gundlach and  M. Ortiz, Phys. Rev. {\bf D 42}, 2521 (1990). 

\noindent 13. B. Boisseau and  B. Linet, Gen. Rel. Grav. {\bf 30}, 963 (1998).

\noindent 14. S. Weinberg, {\it Gravitation and Cosmology}, John Wiley and sons, New York (1972).

\noindent 15. C. W. Misner, K. S. Thorne and J. Wheeler, {\it Gravitation}, W. H. Freeman and company, New York (1997).

\noindent 16. A. Barros and C. Romero, J. Math. Phys. {\bf 36}, 5800 (1995).

\end{document}